\documentstyle[11pt]{article}

\begin{document}
\input{PSMACROS.TEX}
\large

\def\lsim{\mathrel{\rlap{\lower3pt\hbox{\hskip0pt$\sim$}}
    \raise1pt\hbox{$<$}}}         
\def\gsim{\mathrel{\rlap{\lower4pt\hbox{\hskip1pt$\sim$}}
    \raise1pt\hbox{$>$}}}         
\def\dblint{\mathop{\rlap{\hbox{$\displaystyle\!\int\!\!\!\!\!\int$}}
    \hbox{$\bigcirc$}}}
\def\ut#1{$\underline{\smash{\vphantom{y}\hbox{#1}}}$}

\newcommand{\ind}[1]{_{\begin{small}\mbox{#1}\end{small}}}
\newcommand{\WA}{{\em WA}}
\newcommand{\SM}{Standard Model }
\newcommand{\QCD}{{\em QCD }}
\newcommand{\KM}{{\em KM }}
\newcommand{\hscale}{\mu\ind{hadr}}
\newcommand{\aver}[1]{\langle #1\rangle}

\newcommand{\appa}{\mbox{\ae}}
\newcommand{\CP}{{\em CP}}
\newcommand{\fy}{\varphi}
\newcommand{\hi}{\chi}
\newcommand{\al}{\alpha}
\newcommand{\as}{\alpha_s}
\newcommand{\gf}{\gamma_5}
\newcommand{\gm}{\gamma_\mu}
\newcommand{\gn}{\gamma_\nu}
\newcommand{\be}{\beta}
\newcommand{\ga}{\gamma}
\newcommand{\de}{\delta}
\renewcommand{\Im}{\mbox{Im}}
\renewcommand{\Re}{\mbox{Re}}
\newcommand{\GeV}{\,\mbox{GeV }}
\newcommand{\MeV}{\,\mbox{MeV }}
\newcommand{\matel}[3]{\langle #1|#2|#3\rangle}
\newcommand{\state}[1]{|#1\rangle}
\newcommand{\ra}{\rightarrow}
\newcommand{\ve}[1]{\vec{\bf #1}}

\vspace*{.4cm}
\begin{flushright}
\large{
CERN-TH.7050/93\\
UND-HEP-93-BIG\hspace*{0.1em}04}\\
October 1993\\
\end{flushright}
\vspace{0.4cm}
\begin{center} \LARGE {
MARCHING TOWARDS A THEORY OF NON-LEPTONIC DECAYS OF
HEAVY-FLAVOUR HADRONS \footnote{Invited lecture given at the
Advanced Study Conference on Heavy Flavours, Pavia, Italy,
Sept. 1993}}
\end{center}
\vspace{.4cm}
\begin{center} \Large I.I. Bigi
\\
{\normalsize{\it Theoretical Physics Division, CERN\\
CH-1211 Geneva 23, Switzerland
\footnote{During the academic year 1993/94}}\\
and\\
{\it Dept. of Physics,
University of Notre Dame du
Lac, Notre Dame, IN 46556, U.S.A.
\footnote{Permanent address}
\\e-mail address: VXCERN::IBIGI, BIGI@UNDHEP}}

\vspace{.4cm}

\end{center}
\thispagestyle{empty} \vspace{.4cm}

\centerline{\Large\bf Abstract}
\vspace{.4cm}
Theoretical descriptions for the nonleptonic decays of
heavy flavour hadrons have emerged that incorporate
non-perturbative corrections in a systematic way and
are genuinely based on QCD. One can reproduce
$\tau (D^+)/\tau (D^0)$, $BR_{SL}(D^+)$ and
$BR_{SL}(D^0)$ as due mainly to destructive interference
in $D^+$ decays; one predicts $\tau (B^-)$ to exceed
$\tau (B_d)$ by a few per cent; $BR_{SL}(B)$ is
reduced only slightly, raising the prospects of a
serious and tantalizing conflict with ARGUS \&
CLEO data. It is emphasized that the lifetimes for
the two mass eigenstates in the $B_s$-$\bar B_s$ system
could differ by 30\% or even more.
The status of phenomenological treatments of
non-leptonic two-body modes is briefly reviewed.

\vspace{.4cm}

\large
\addtocounter{footnote}{-3}

\section{Introduction}

While the top quark, yet to be discovered, represents a
heavy flavour {\it par excellence}, I will not address it in
my talk here. For its decay rate into an {\em on-shell} W boson
will in all likelihood be so rapid that it
cannot form a hadron\cite{RAPALLO}; instead it will decay truly as a quark,
affected by perturbative QCD only, without the usual
complications of non-leptonic decays. It is therefore
the decays of charm and beauty hadrons
that I will discuss.
I will actually focus on two complementary
classes of transitions, namely inclusive decays on the one hand
-- total lifetimes, semileptonic branching ratios,
lepton spectra -- and exclusive non-leptonic two-body modes
on the other.
For one encounters considerably larger theoretical
uncertainties in other decay classes, such as genuine
three-body modes etc.

The talk will be organized as follows:
in Sect. 2 I treat inclusive transitions whereas
in Sect. 3 I discuss non-leptonic two-body modes, before
presenting a summary and an outlook in Sect.4.

\section{Inclusive Decays of Beauty and Charm}

In discussing total rates and the lepton spectra in
semileptonic decays, I will mainly follow the analysis
given by Uraltsev, Vainshtein, Shifman, Blok and
myself \cite {US,SPECTRA,THEM}.
\subsection{Total Rates}
\subsubsection{Formalism}
The basic procedure can be illustrated by a simple analogy
with nuclear $\beta$ decay
due to Shifman. There are two
effects distinguishing the decay of a neutron bound
inside a nucleus from that of a free neutron:

(a) nuclear binding effects;

(b) Pauli statistics correlating the electrons surrounding
the nucleus with those emerging from the $\beta$
decay.

The typical energies of the bound electrons are
certainly small
compared to $E_{rel}$, the energy released in the decay;
let us assume -- although this is not true in
reality -- that also the nuclear binding energies were
small compared to $E_{rel}$. In that case nuclear $\beta$
decays proceed like the decays of {\em free}
neutrons to a good approximation; corrections to this
simple `spectator' picture can be computed via an expansion in
powers of $1/E_{rel}$. In practice no use is made of this
observation for nuclear $\beta$ decay since these corrections
can be incorporated in a more compact form by explicitly
using the wavefunctions of the bound nucleons and
electrons.

There arise analogous corrections to the decay rate
for a quark $Q$ inside a hadron:

(a) interactions of the decaying quark with other partons in
the hadron, due to the exchange of gluons or
of W bosons;

(b) Pauli interference effects of the decay products with
other partons in the hadron.

Yet the difference is that even the
concept of a (hadronic) wavefunction is of dubious value. The
most reliable approach is then to evaluate weak decay rates
of heavy flavour hadrons through an expansion in inverse
powers of the heavy flavour quark mass $m_Q$. The width for
the decay of a heavy flavour hadron $H_Q$ into an inclusive
final state $f$ is expressed as follows:
$$\Gamma (H_Q\ra f)=\frac{G_F^2m_Q^5}{192\pi ^3}|KM|^2
[ c_3(f)\matel{H_Q}{\bar QQ}{H_Q}+
c_5(f)\frac{\matel{H_Q}{\bar Qi\sigma \cdot GQ}{H_Q}}
{m_Q^2}+$$
$$ +\sum _i c_6^{(i)}(f)\frac{\matel{H_Q}
{(\bar Q\Gamma _iq)(\bar q\Gamma _iQ)}{H_Q}}
{m_Q^3} + {\cal O}(1/m_Q^4)],  \eqno (1)$$
where the dimensionless coefficients $c_i(f)$ depend on the
parton level characteristics of $f$ and on the ratios of the
final state quark masses to $m_Q$; $KM$ denotes the
appropriate combination of weak mixing angles.

It is through the expectation values of the operators
appearing on the right-hand side of eq. (1)
that the dependence on the decaying $hadron$, and on
non-perturbative forces in general, enters rather than
through wavefunctions as in nuclear $\beta$ decay.
Since these are on-shell hadronic matrix elements one
sees that $\Gamma (H_Q\ra f)$ is indeed expanded into
a power series in $\mu _{had}/m_Q$ with
$\mu _{had}\, <\, 1\, \mbox{GeV}$. This parameter is quite
small for $Q=b$; non-perturbative corrections are thus
expected to be smallish in beauty decays with the first few
terms in the expansion yielding a good approximation to the
exact result. For $Q=c$ one has an expansion parameter that
-- while smaller than one -- is not really small. {\it A priori}
one can therefore expect to make only semiquantitative
predictions for charm decays.

One can immediately read off an important qualitative result
from eq. (1): to order $1/m_Q$ there are no corrections to
$total$ rates [although they emerge for lepton $spectra$
\cite {SPECTRA,THEM}].
The leading non-perturbative corrections thus
scale like $1/m_Q^2$, i.e. they fade away quickly with
$m_Q$ increasing.

It will be of use here to compare
our approach to
what is usually referred to as Heavy Quark Symmetry or
Effective Heavy Quark Theory ($EHQT$) \cite {EHQT}
in dealing with
$Q_1\ra Q_2`W',\; `W'\ra q'\bar q,\; l\nu $:

-- $EHQT$ applies to $exclusive$ semileptonic decays; it yields
absolute predictions as long as both the initial and final
state quarks are heavy, i.e. for $m_{Q_1},m_{Q_2}\gg
\mu _{had}$, as is approximately
the case for $b\ra cl\nu$, but not
for $b\ra ul\nu$.

-- Our treatment applies to both non-leptonic and
semileptonic $inclusive$ transitions; it requires, as already
stated, the energy release to be large, i.e.
$m_{Q_1}-m_{Q_2}\gg \mu _{had}$, which holds for
$b\ra c$ as well as for $b\ra u$. On the other hand
$b\ra c'$ could not be treated if $m_b\sim m_{c'}$
and neither can $b\ra c\tau \nu$;
on the other hand, $EHQT$ still applies.

Thus there is no conflict between
these two approaches -- they actually
complement each other quite nicely.

At present one is unable to determine the size of the
matrix elements from first principles; yet we can
relate them to other observables. The expectation value of the
chromomagnetic operator $\bar Qi\sigma \cdot GQ$ can be
extracted from the measured hyperfine splitting between the
vector and pseudoscalar mesons:
$$\matel{P_Q}{\bar Qi\sigma \cdot GQ}{P_Q}
\simeq \frac {3}{2} (M_{V_Q}^2-M_{P_Q}^2)\eqno(2)$$
with $V_Q=D^*[B^*]$ and $P_Q=D[B]$ for $Q=c[b]$. One
finds:
$$G_c\equiv \matel{D}{\bar ci\sigma \cdot Gc}{D}/
m_c^2\simeq 0.32, \; \;
G_b\equiv \matel{B}{\bar bi\sigma \cdot Gb}{B}/
m_b^2\simeq 0.03\; .\eqno(3)$$
The expectation value vanishes for baryons $\Lambda _Q$:
$$\matel{\Lambda _Q}{\bar Qi\sigma \cdot GQ}{\Lambda _Q}
\simeq 0\; .\eqno(4)$$
The operator $\bar QQ$ can be expanded again into a series
of inverse powers of $m_Q$:
$$\matel{H_Q}{\bar QQ}{H_Q} = 1-\frac{\matel{H_Q}{(\vec p)^2}
{H_Q}}{2m_Q^2}+\frac{3}{8}\cdot \frac{M_{V_Q}^2-M_{P_Q}^2}
{m_Q^2}+{\cal O}(1/m_Q^3)\; ,\eqno(5)$$
where $\matel{H_Q}{(\vec p)^2}{H_Q}\equiv
\matel{H_Q}{\bar Q(i\vec D)^2Q}{H_Q}/2m_Q$ denotes the
expectation value for the kinetic energy of the
heavy quark $Q$
moving inside the hadron $H_Q$ under the influence of the
gluon background field. The first term on the right-hand side
of eq. (5) reproduces the simple parton model result, i.e.
the `spectator ansatz' yielding universal lifetimes
and semileptonic branching ratios for all hadrons $H_Q$;
it indeed dominates for $m_Q\ra \infty$.
The first two terms represent the mean value
of the Lorentz time dilatation factor $\sqrt(1-v^2)$
slowing down the decay of the quark $Q$ in a moving frame.
The numerical size of
$\matel{H_Q}{(\vec p)^2}{H_Q}$ is not well known
yet. Results from two analyses based on QCD sum rules
exist though\cite {NEUBERT};
lattice simulations of QCD will be able
to extract this quantity in the near future for charm
\footnote {Lattice computations will actually yield
primarily $\matel {H_c}{\bar cc}{H_c}$.}; measuring the mass
of $\Lambda _b$ to 10 MeV accuracy will enable us to determine
the difference $\matel{\Lambda _Q}{(\vec p)^2}{\Lambda _Q}-
\matel{P_Q}{(\vec p)^2}{P_Q}$. At present we have to content
ourselves with the rough estimate
$\matel{H_Q}{(\vec p)^2}{H_Q}
\sim 0.5\; (\mbox{GeV})^2$ and thus
$$K_c\equiv \matel {H_c}{\bar c(i\vec D^2c}{H_c}/2m_c^2
\sim 0.1,\; \; \;
K_b\equiv \matel {H_b}{\bar b(i\vec D^2b}{H_b}/2m_b^2
\sim 0.01\; .\eqno(6)$$
One sees from comparing eqs. (3) and (4) that the leading
non-perturbative corrections arising on the $1/m_Q^2$
level differentiate between meson decays on the one
hand and baryon decays on the other.

Differences between meson lifetimes and semileptonic
branching ratios arise in order $1/m_Q^3$. It has been
shown \cite {WA} that the corrections due to
`Pauli Interference' and `Weak Annihilation/ Weak
Scattering' can be expressed through local matrix elements
of current-current operators --
$\matel{H_Q}{\bar Q\Gamma _iq\bar q\Gamma _iQ}{H_Q}
\simeq \frac{4}{3} f_Q^2m_Q$ with $f_Q=f_D,\; f_B$
for the mesons
and similarly for the baryons. Thus one finds
$$\frac{\Gamma _{nonspect}}{\Gamma _{spect}}\propto
\frac{f_Q^2}{m_Q^2}\; .\eqno(7)$$

\subsubsection{Some Results}

{\bf (A)}
To order $1/m_Q^2$ one obtains for the semileptonic and the
non-leptonic widths:
$$\Gamma _{SL}(H_Q\ra q)=\frac{G_F^2m_Q^5|V_{Qq}|^2}
{192\pi ^3}(z_0\matel{H_Q}{\bar QQ}{H_Q}-
\frac{z_1}{m_Q^2}\matel{H_Q}{\bar Qi\sigma \cdot GQ}{H_Q})
\eqno(8)$$
$$\Gamma _{NL}(H_Q\ra q)=\frac{G_F^2m_Q^5|V_{Qq}|^2N_C}
{192\pi ^3}\cdot $$
$$\cdot [A_0(z_0\matel{H_Q}{\bar QQ}{H_Q}-
\frac{z_1}{m_Q^2}\matel{H_Q}{\bar Qi\sigma \cdot GQ}{H_Q})
-\frac{4A_2z_2}{m_Q^2}
\matel{H_Q}{\bar Qi\sigma \cdot GQ}{H_Q}]\; ,
\eqno(9)$$
with $z_0=1-8x+8x^3-x^4-12x^2\log (x)$, $z_1=(1-x)^4$,
$z_2=(1-x)^3$ and $x=m_q^2/m_Q^2$, describing the relative
phase spaces and $A_0=(c_+^2+c_-^2)/2+(c_+^2-c_-^2)/2N_C$,
$A_2=(c_+^2-c_-^2)/2N_C$
denoting the radiative QCD corrections
($N_C$ = number of colours).

There are two observations that should be made here:

\noindent (i) The semileptonic branching ratios for
heavy flavour $mesons$ are $reduced$
with respect to the parton model expectation due
to the last term in eq. (9)
since $A_2<0$. Numerically one finds for $B$ decays
\footnote {The size of this correction (although not its
sign) can be anticipated by noting that it is
an effect of order $(\mu _{had}/m_b)^2 \sim
{\cal O}(\%)$ for $\mu _{had}\leq 1\; \mbox{GeV}$.}
$$\frac{\delta BR_{SL}(B)}{BR_{SL}(B)}\simeq
6\cdot \frac{A_2z_2}{A_0z_0}\cdot
\frac{M_{B^*}^2-M_B^2}{m_b^2}\cdot BR_{NL}(B)
\sim - 0.05\; .\eqno(10)$$
Next-to-leading order perturbative corrections, which have not
been included here, could reduce it further,
but probably not below 12\%.
A much larger correction arises in $D$ decays, reflecting
the scaling law $(m_b/m_c)^2$:
$$\frac{\delta BR_{SL}(D)}{BR_{SL}(D)}
\sim -{\cal O}(50\%)\eqno(11)$$
i.e. $BR_{SL}(D) \sim 10\%$
instead of $BR(c\ra l\nu s)\simeq 15\%$.

\noindent (ii) The semileptonic $widths$ of
$\Lambda _c\; [\Lambda _b]$ and $D\; [B]$ differ
by roughly 30\% [a few \%].

\noindent {\bf (B)} Lifetime differences
among mesons arise on the
$1/m_Q^3$ level:
$$\frac{\tau (D^+)}{\tau (D^0)}\sim 2,\; \; \; \;
\frac{\tau (D_s)}{\tau (D^0)}\simeq 1\; ,\eqno(12)$$
$$\frac{\tau (B^-)}{\tau (B_d)}\simeq
1+0.05\cdot \frac{f_B^2}{(200\; \mbox{MeV})^2}\; ,\eqno(13)$$
$$\frac{\tau (\Lambda _b)}{\tau (B_d)}
\sim 0.85-0.9\; .\eqno(14)$$
Eq. (14) represents a ballpark estimate only since no
detailed analysis has been done yet.

It is at least amusing to note that the largest lifetime
difference among $B$ mesons is produced by a
subtle mechanism, namely $B_s$-$\bar B_s$ oscillations:
$$\frac{\Delta \Gamma (B_s)}{\bar \Gamma (B_s)}
\equiv \frac{\Gamma (B_{s,short})-\Gamma (B_{s,long})}
{\bar \Gamma (B_s)}\simeq 0.18\cdot
\frac{f_{B_s}^2}{(200\; \mbox{MeV})^2}\; .\eqno(15)$$
One can search for the existence of two
different $B_s$ lifetimes by
comparing $\tau (B_s)$ as measured in
$B_s\ra \psi \phi$ and in $B_s\ra l\nu X$:
$|\Gamma (B_s\ra \psi \phi)-\Gamma (B_s\ra l\nu X)|
\simeq \frac{1}{2}\Delta \Gamma (B_s)$.
\footnote {Analogously one can compare
$\tau (B_d)$ as obtained from $B_d\ra \psi K_S$, from
$B_d\ra \psi K^*$ and from $B_d\ra l\nu X$. Yet theoretically
one expects a lifetime difference on the per cent level only.}
Whether an
effect of the size indicated in eq. (15) is large
enough to be ever observed in a real experiment is of
course a different matter. It has to be said, though,
that eq.(15) does not represent a `gold-plated' prediction.
It is conceivable that the underlying computation,
which invokes quark box diagrams, underestimates the actual
lifetime difference.

\subsection{Lepton Spectra}

The method outlined above can be extended to treat the
lepton energy spectra in $H_Q\ra l\nu X$ transitions. The
expansion is now in $1/(1-y)m_Q$ rather
than $1/m_Q$, where $y=2E_l/m_Q$ denotes the normalized
lepton energy. This expansion is obviously singular at
$y=1$, i.e. in the endpoint region, and one has to
apply some care in interpreting the results there.

The spectrum
$d\Gamma /dy$ is evaluated in principle without any free
parameters, although in practice there is at present some
numerical uncertainty in the size of $K_Q$, as stated before.
One obtains the following results \cite{SPECTRA}:

$\bullet$ The shape of the spectrum is
remarkably similar to the
phenomenological model of Altarelli et al.
\cite{ALT}
that has been
fitted to the data; there are $\sim 10 \%$ differences
in normalization.

$\bullet$ Differences in the two descriptions do occur in
the endpoint region.

$\bullet$ For consistency reasons one has to use
the current mass
$m_u\simeq 0$ in the QCD treatment of the $b\ra u$
spectrum. It turns out that this spectrum is very similar
to the one obtained in a parton model description with
$m_u\sim 300\; \mbox{MeV}$; i.e. the non-perturbative
corrections produce effectively a constituent mass for the
light quark.

$\bullet$ The endpoint spectrum will look quite
different for $B_d$ and $B^-$ decays. Likewise
for the lepton spectra in $D^0$ vs. $D^+$ vs. $D_s$
decays \cite{WAII}.

\section{Exclusive Two-Body Decays of Beauty and Charm}

A discussion of exclusive non-leptonic decays has to be opened
with a note of caution:

{\em First Theoretical Caveat: The relationship between
inclusive and exclusive transition rates is
nothing short of delicate!}

This piece of common sense can be illustrated by the following
example. Consider the {\em corrections} to the decay width of a
$(Q\bar q)$ meson that are induced by Weak Annihilation, see
fig.~\ref{FIGURE}.
\psfig{BIGI_FIGURE.EPS}{70}{70}
{Corrections to the $Q\bar q \ra Q\bar q$
forward scattering
amplitude induced by Weak Annihilation. The three cuts
(a), (b) and (c) represent different
final states for the $Q\bar q$ decay.}{FIGURE}{10}{10}{30}
As indicated there are three cuts in the
$Q\bar q \ra Q\bar q$ forward scattering amplitude
representing different final states, one with an
on-shell gluon and the other two with a (slightly)
off-shell gluon. Summing over these three cuts yields an
overall correction that remains finite even in the limit
$m_q\ra 0$ \cite{WA}.
However a very striking pattern emerges when one
considers separately the three `exclusive' channels $(a)$,
$(b)$ and $(c)$
\footnote{In the real world these three channels can of course
not be distinguished; yet this academic model can illustrate the
relevant point.}:
The contribution $(b)$ constituting the square of an
amplitude is positive; in the limit $m_q\ra 0$ it is dominated
by a term $+\frac{1}{m_q^2}|T|^2$; the contributions $(a)$
and
$(c)$ on the other hand represent interference terms that,
taken together, are of
the form $-\frac{1}{m_q^2}|T|^2$ for $m_q\ra 0$ since the
sum of $(a)$, $(b)$ and $(c)$
has to possess a regular limit.
I want to draw the following lesson from this discussion: a
small effect in an overall rate can be due to large
cancellations among subclasses of decays. The relevance of this
statement will become clearer later on.

\subsection{Phenomenological Models}

Phenomenological descriptions of non-leptonic two-body decays
of charm and beauty were pioneered by the authors of
ref. \cite{BSW}.
There are three main ingredients in such models: \break
(i) One assumes
factorization, i.e. one uses $\matel {M_1M_2}{J_{\mu}J_{\mu}}
{D} \simeq \matel{M_1}{J_{\mu}}{D}\cdot \matel{M_2}{J_{\mu}}
{0}$ to describe $D\ra M_1M_2$. (ii) One employs one's
favourite hadronic wavefunctions to compute
$\matel{M_1}{J_{\mu}}{D}$. Very recently Heavy Quark Symmetry
and Chiral Symmetry (for the
light quarks) have been incorporated into these wavefunctions
\cite{GATTO}.
(iii) All two-body modes are then
expressed in terms of two free fit parameters $a_1^{(c)}$
and $a_2^{(c)}$,
with $a_1^{(c)}$ controlling the `class I'
$D\ra M_1^+M_2^-$
and $a_2^{(c)}$ the `class II' $D\ra M_1^0M_2^0$ transitions;
both quantities contribute coherently to the `class III'
transitions $D^+\ra M_1^0M_2^+$. The analogous procedure is
followed for $B$ decays allowing though for different values
for $a_1^{(b)}$ and $a_2^{(b)}$.

With these two free parameters $a_{1,2}^{(c)}$ (and some
considerable degree of poetic license in invoking strong
final state interactions) one obtains a decent fit for the
$D^0$ and $D^+$ modes (much less so however for $D_s$
decays). The situation is similar in $B$ decays. One has to
point out, however, that this success is helped considerably
by the forgiving imprecision in many of the branching ratios
measured so far.

A more ambitious program had been put forward, where one undertakes
to infer $a_{1,2}$ from QCD: the effective transition operators
are renormalized with coefficients $c_+,c_-$ generated
from perturbative QCD; in evaluating hadronic matrix elements
one retains -- and that is an ad-hoc assumption --
only the terms leading in $1/N_C$:
$a_1=(c_++c_-)/2,\; a_2=(c_+-c_-)/2$.
Since the short-distance coefficients depend logarithmically
on the heavy flavour mass, it then follows that
$a_1^{(c)}\sim a_1^{(b)}$ and $a_2^{(c)}\sim a_2^{(b)}$.
At this point I would like to give two more notes of caution:

{\em Second Theoretical Caveat: There exists no
general proof of the Dogma of Factorization for real hadrons; I
actually
consider it unlikely to be of universal validity,
in particular
for class II transitions.}
It thus makes eminent sense to subject this dogma to as many
different experimental tests as possible \cite {HON} .

{\em Third Theoretical Caveat: It is very unlikely that the
rule of retaining only leading terms in $1/N_C$ is
universally
implemented in QCD.} Our analysis of inclusive heavy-flavour
decays actually found cases where this rule was dynamically
realized (i) for $D$ as well as for $B$ decays,
(ii) for $D$, but not for $B$ decays, or (iii) for neither.

I am quite certain that more precise data will reveal more
and more systematic deficiencies in these models. They are already
emerging in the present data: as mentioned before
strong final state interactions have to be invoked as
`deus ex machina';
the branching ratio for the channel $D^+\ra \pi ^0\pi ^+$,
where no such rescue is possible, appears to be seriously
underestimated \cite{ARTUSO}.
Yet the most intriguing surprise has
surfaced in $B$ decays: CLEO data now show with
large statistical significance that class III
transitions like $B^-\ra D^0\pi ^-$ proceed considerably
faster than the class I modes $B_d\ra D^+\pi ^-$.
\cite{HON}.
This means that $a_1^{(b)}\cdot a_2^{(b)} > 0$ holds, i.e.
that $constructive$ interference takes place in $B$ decays.
This is surprising in three respects: (i) It is in clear
contrast to the situation in $D$ decays. (ii) While it does
not pose any fundamental problem for the BSW model, it
represents a basic failing for the $1/N_C$ ansatz. (iii) It
raises the question of whether the same constructive interference
might occur for the {\em inclusive} rate thus shortening
$\tau (B^-)$ relative to $\tau (B_d)$ rather than
lengthening it.

As stated in Sect. 2 the $B^-$ lifetime is predicted to
exceed the $B_d$ lifetime by a few per cent only; yet one has
to keep in mind the {\em First Theoretical Caveat} stated in the
beginning of this section: a small correction in the inclusive
rate is quite likely to be made up by large contributions
of alternating signs coming from different
classes of exclusive transitions.

For a better understanding of these problems one has to
progress to treatments that are rooted more firmly in QCD.
However I would like to first stress the important lessons
we have learnt and are still learning from these
phenomenological descriptions:

-- They have yielded quite a few successful predictions in a
`user-friendly' way.

-- They have helped us considerably in focusing on the underlying
theoretical problems such as the question of factorization or the
$1/N_C$ rule.

-- From their fits to the data they provide us with valuable,
albeit indirect information on the final-state interactions.
Such information is crucial in studies of direct CP violation.

\subsection{Theoretical Analysis}

The first treatment of two-body decays of $D$ mesons that is
intrinsically connected to QCD was given by Blok and Shifman
some time ago, based on QCD sum rules
\cite{BLOKI}. It would be desirable to
have this analysis updated and refined for $D$ decays and extended
to $B$ decays.

The same two authors have moved in a different direction, namely to
apply methods based on heavy quark expansions to non-leptonic
two-body modes of beauty. They argue that the $1/N_C$ rule
is dynamically implemented in the class I $B_d\ra D^+\pi ^-$
mode whereas the situation is considerably more complex in
class II and III transitions
\cite{BLOKII}. I believe this ansatz promises
to advance our theoretical understanding of exclusive heavy flavour
decays and thus deserves increased attention.

Lattice simulations of QCD are reaching a level where they can
treat $D$ decays. This will be covered by other speakers at this
conference.

\subsection{Prizes to be Attained}

The theoretical methods one applies to exclusive decays are
often not of the most lucid kind. Yet they
are essential (if imperfect) tools for addressing fundamental
questions. Let me just cite a few topical examples:

\noindent (i) $D^0$-$\bar D^0$ oscillations in the
Standard Model are in all likelihood dominated by
virtual transitions into two-body modes common to
$D^0$ and $\bar D^0$ mesons:
$$D^0\ra \pi \pi, K\bar K, K\pi,
\bar K\pi \ra \bar D^0 \; .$$
Understanding them would allow us to
say whether a signal for $D^0$-$\bar D^0$ oscillations that
might be observed in the future requires the intervention
of `New Physics' or is still compatible with the Standard
Model.

\noindent (ii) $\Delta \Gamma (B_s)$, i.e. the lifetime
difference between $B_{s,short}$ and $B_{s,long}$,
is usually computed from the quark box diagram with internal
$c$ (and $u$) quarks, leading to a result like
the one quoted in eq. (15).
However the weight of such a short-distance contribution to
$\Delta \Gamma (B_s)$ is much more uncertain than that of the
local contribution to $\Delta m(B_s)$, which is given
by virtual top exchanges. It is therefore
conceivable that long-distance dynamics could provide a large
or even dominant
contribution through
the transitions
$$B_s\ra D_s^{(*)}\bar D_s^{(*)}, \psi \phi, \psi \eta
\ra \bar B_s\; ,$$
This has been analysed by the Orsay group \cite {ORSAY},
which infers from present data $\Delta \Gamma (B_s)\simeq
0.15$.

\noindent (iii) The presence of non-trivial strong final state
interactions is neccessary for direct CP violation to become
observable. Prof. Artuso mentioned some good news in that
respect at this conference when she discussed the observed
$SU(3)_{fl}$ breaking in Cabibbo-suppressed $D^0$
decays \cite{ARTUSO}:
$$\frac{\Gamma (D^0\ra \pi ^+ \pi ^-)}
{\Gamma (D^0\ra K^+ K^-)}=0.38 \pm 0.15,\; \; \;
\frac{\Gamma (D^0\ra \pi ^+ \pi ^-,\pi ^0\pi ^0)}
{\Gamma (D^0\ra K^+ K^-, K^0\bar K^0)}=0.37 \pm 0.15\; .
\eqno(16)$$
Assuming factorization, one would expect a considerably higher
number, namely $\sim (f_{\pi}/f_K)^2 \simeq 0.7$ for
these ratios, in particular for the second one in eq. (16)
since it is not sensitive to charge-exchange processes in the
final state. Factorization thus provides a poor description
and nontrivial final state interactions play a major role in
these modes \cite{KAMAL}.
A nice at least qualitative description of this
situation has been given a long time ago by Fukugita, Hagiwara
and Sanda \cite{SANDA}, who pointed out that
(non-local) penguin operators
contribute {\em destructively} to $D^0\ra \pi \pi$ and
{\em constructively} to $D^0\ra K^+ K^-$.

\section{Summary and Outlook}
\subsection{Status}
New and more powerful second-generation
theoretical technologies are
emerging: QCD sum rules, Heavy Quark Symmetry, $1/m_Q$
expansions and lattice simulations of QCD. They are leading
to

$\bullet$ significant conceptual progress, namely a better
understanding of (i) the form and size of non-perturbative
corrections,
(ii) the relationship between charm and
beauty decays, where the former play the role of a microscope
for the non-perturbative corrections in the latter,
and (iii) the differences and similarities of baryon vs. meson
decays;

$\bullet$ the realization that charm and beauty baryons deserve
detailed studies in their own right.

$\bullet$ a quantitative phenomenology that is genuinely based
on QCD:
$$\frac {\tau (D^+)}{\tau (D^0)}\sim 2; \; \;
BR_{SL}(D^+)\sim 16\%,\; \; BR_{SL}(D^0)\sim 8\%;
\; \; \frac{\tau (D_s)}{\tau (D^0)}\sim 1.0\pm {\cal O}(\%)
\; ,\eqno(17)$$
i.e. the data are reproduced within the accuracy of the
expansion.
$$d\Gamma (D_s\ra lX)/dy\neq d\Gamma (D^0\ra lX)/dy
\neq d\Gamma (D^+\ra lX)/dy\eqno(18)$$
$$\frac{\tau (B^-)}{\tau (B_d)}\simeq
1+0.05\cdot \frac{f_B^2}{(200\; \mbox{MeV})^2}; \; \; ;
BR_{SL}(B)\simeq
12-13\% \eqno(19)$$
$$\frac{\Delta \Gamma (B_s)}{\bar \Gamma (B_s)}
\simeq 0.18\cdot \frac{f_{B_s}^2}{(200\; \mbox{MeV})^2},\;\; \; \;
\frac{\tau (\Lambda _b)}{\tau (B_d)}
\sim 0.85-0.9\eqno(20)$$
$$d\Gamma (B^-\ra lX)/dy\neq d\Gamma (B_d\ra lX)/dy
\eqno(21)$$
The prediction for $BR_{SL}(B)$ is somewhat larger than
present CLEO and ARGUS measurements. This could turn
out to be a
serious -- or intriguing -- discrepancy. It could
conceivably signal the presence of anomalously
large higher-order contributions that so far have not
been included in the
theoretical expression. In that case one would expect
lifetime ratios for beauty hadrons to differ more from
unity than stated in eqs. (19) and (20)\cite{PUZZLE}.

\noindent Phenomenological models for nonleptonic two-body modes
are encountering discrepancies with more precise
data; yet they
continue
to be useful and help us in focusing on the underlying theoretical
issues.

\subsection{Future}
One can expect a refinement of and
increased cooperation (rather than just coexistence) between
the second-generation theoretical technologies.
On the experimental side one can hope for

-- lifetime measurements for individual beauty hadrons with
$\pm 10\%$ accuracy soon and $\pm \%$ in the longer run;

-- data on $\tau (B_s)$ {\em separately} from
$B_s\ra \psi \phi$ and from $B_s\ra l\nu D_s$;

-- measurements of {\em absolute} branching ratios
for $D_s$, $\Lambda _c$ and $\Xi _c$, in particular their
semileptonic branching ratios;

-- perform the `class I, II, III' phenomenology
{\em individually} for {\it KM} allowed and {\it KM}
suppressed $B$  decays.

The primary goal in all these efforts is to be able
to exploit to the fullest over
the next 20 years or so, the
discovery potential or even discovery guarantee
that awaits us in beauty physics. It certainly would be
a crime not to make all conceivable efforts to obtain the
required experimental facilities. Yet this does not mean
that the chapter on charm decays should be
closed. For those represent
the best laboratory to test and
sharpen our QCD tools as they are relevant
for the decays of heavy-flavour hadrons. This can be
done optimally at a $\tau$-charm factory. Furthermore we
cannot afford to overlook the possibility that `New Physics'
might surface in unexpected areas, namely in charm decays!

\vspace*{0.5cm}

{\bf Acknowledgements:} \hspace{.4em} I gratefully acknowledge
many illuminating discussions on the
subject of this paper with
several colleagues, and in particular with N. Uraltsev,
M. Shifman, B. Blok and A. Vainshtein.
This work was supported by the National
Science Foundation under grant number PHY 92-13313.

\vspace*{1cm}

\vspace*{0.5cm} \\

%
%
%
%
%
%
\newlength{\figlth}
\newlength{\fighof}
\newlength{\figvof}
\def\psfig#1#2#3#4#5#6#7#8{
\begin{figure}[htbp]
   \setlength{\figlth}{#8in}
   \setlength{\fighof}{63pt}
   \setlength{\figvof}{7.2pt}
     \divide \figlth by 1000
     \multiply \figlth by #3
     \divide \fighof by 2
     \advance \fighof by #6pt
     \divide \fighof by 1000
     \multiply \fighof by #2
     \advance \fighof by -2.8pt
     \multiply \fighof by -1
     \multiply \fighof by 72
     \divide \figvof by 100
     \multiply \figvof by #7
     \multiply \figvof by #3
     \multiply \figvof by -1
\vspace{\figlth}
\includegraphics{#1}
\caption{ #4 \label{#5}}

\end{figure}}
\end{document}